\documentclass[10pt,letterpaper]{article}
\usepackage[top=0.85in,left=2.75in,footskip=0.75in]{geometry}

\usepackage{amsmath,amssymb}

\usepackage{changepage}

\usepackage[utf8x]{inputenc}

\usepackage{textcomp,marvosym}

\usepackage{cite}
\usepackage{color}
\usepackage{nameref,hyperref}

\usepackage[right]{lineno}

\usepackage{microtype}
\DisableLigatures[f]{encoding = *, family = * }

\usepackage[table]{xcolor}

\usepackage{array}

\newcolumntype{+}{!{\vrule width 2pt}}

\newlength\savedwidth

\raggedright
\usepackage[aboveskip=1pt,labelfont=bf,labelsep=period,justification=raggedright,singlelinecheck=off]{caption}

\makeatletter
\renewcommand{\@biblabel}[1]{\quad#1.}
\makeatother

\date{}

\usepackage{lastpage,fancyhdr,graphicx}
\usepackage{epstopdf}
\fancyheadoffset[L]{2.25in}
\fancyfootoffset[L]{2.25in}

\begin{document}
\vspace*{0.2in}

\begin{flushleft}
{\Large
\textbf\newline{Subthreshold signal encoding in coupled FitzHugh-Nagumo neurons} 
}
\newline
\\
Maria Masoliver\textsuperscript{1},
Cristina Masoller\textsuperscript{1*}

\bigskip

\textbf{1} Departament de Fisica, Universitat Politecnica de Catalunya, Rambla de Sant Nebridi, 22, 08222 Terrassa, Barcelona, Spain

* cristina.masoller@upc.edu

\end{flushleft}
\section*{Abstract}
Despite intensive research, the mechanisms underlying how neurons encode external inputs remain poorly understood. Recent work has focused on the response of a single neuron to a weak, subthreshold periodic signal. By simulating the FitzHugh-Nagumo stochastic model and then using a symbolic method to analyze the firing activity of the neuron, preferred and infrequent spike patterns (defined by the relative timing of the spikes) were detected, whose probabilities encode information about the signal. As not individual neurons in isolation but neuronal populations are responsible for the emergence of complex behaviors, a relevant question is whether this coding mechanism is robust when the neuron is not isolated. We study how a second neuron, which does not perceive the subthreshold signal, affects the detection and the encoding of the signal, done by the first neuron. Through simulations of two coupled FitzHugh-Nagumo neurons we show that the coding mechanism is indeed robust, as the neuron that perceives the signal fires a spike train that has symbolic patterns whose probabilities depend on the features of the signal. Moreover, we show that the second neuron facilitates the detection of the signal, by lowering the firing threshold of the first neuron. This in turn decreases the internal noise level need to fire the spikes that encode the signal. We also show that the probabilities of the symbolic patterns achieve maximum or minimum values when the period of the external signal is close to (or is half of) the mean firing period of the neuron. 

\section*{Author summary}

Neurons encode and transmit information in sequences of spikes, and in spite of intensive research, the principles underlying the neural code are yet not fully understood. In the framework of a simple neuron model, it was recently conjectured that, when a neuron is in a noisy environment and receives a weak periodic input, it encodes the information in the form of preferred and infrequent spike patterns. Here we study how the coupling to a second neuron, which does not receive the external signal, affects the way the first neuron encodes the signal. Our goal is to characterize the role of the second neuron. We show that it has two main effects: first, it decreases the firing threshold, allowing the first neuron to encode the signal at lower noise levels, and second, it modifies the preferred and the infrequent patterns, whose probabilities still encode information about the period and amplitude of the signal. 

\section*{Introduction}

In spite of having been the object of intensive research for decades, the mechanisms used by neuronal populations to encode and transmit information remain poorly understood. Breaking the neural code and yielding light into neuronal strategies for efficient encoding of information in noisy environments is a hot topic in neuroscience research. Advances in this area will not only improve our understanding of brain function, but also, could revolutionize artificial intelligence systems and communication technologies, as new paradigms based on how neurons efficiently encode information could allow to overcome the limitations of present day optical computing systems and communication technologies \cite{optical_neuron_oe_2011,aragoneses_2014,graphene_2016,natphot_2017}.

Various mechanisms have been proposed to explain how neurons encode external inputs, which can been viewed as complementary, or functional, under different situations \cite{thorpe_2001, nature_2002,nature_2003,hidden_2004,coombes_2010}. For example, neuronal populations can encode information in the spike rate, in the spike timing, in the frequency content of spike sequences, in the coherence of spatial spike patterns, etc. Linear and non-linear data-driven methods have been developed to quantify the information content of neuronal activity \cite{eguia_2000,panzeri_nat_rev_2009,ostojic_plos_cb_2011,amigo_2013}. A lot of research has focused on the statistics of the time intervals between consecutive spikes (inter-spike intervals, ISIs) and how properties such as ISI correlations affect information encoding \cite{andre_1991,ratnam_2000,nawrot_2007,nawrot_2009,lindner_plos_comp_bio_2010}.

Recently, the response of an individual neuron to a weak periodic signal was studied numerically \cite{REI16}, in the framework of the FitzHugh-Nagumo model \cite{FIT61a,NAG62}. The analysis focused in a sub-threshold signal, which means that the signal alone does not produce spikes. Therefore, without background noise, the neuron's membrane voltage displays only small, subthreshold oscillations. However, in the presence of noise, the firing activity of the neuron encodes information about  the amplitude and the period of the signal \cite{REI16}. By analyzing the ISI sequence using a nonlinear symbolic method \cite{BAN02}, it was shown that the weak periodic signal induces the emergence of relative temporal ordering in the timing of the spikes, which is absent if the neuron's firing activity is only due to uncorrelated noise \cite{REI16,REI16_2}. Temporal ordering was detected in the form of more and less expressed symbolic patterns, which depend on the period of the signal and on the level of noise. The pattern's probabilities monotonically increase with the amplitude of the signal and thus encode information about both features, the amplitude and the period of the signal.  A resonance-like behavior was found, as certain periods and noise levels enhance temporal ordering, maximizing (or minimizing) the probability of the more (less) expressed pattern. 

An open question is whether this encoding mechanism is robust when a neuron is not in isolation. In particular, can a neuron still use this mechanism to encode a sub-threshold periodic signal, when it is coupled to other neurons that do not perceive the signal? To address this question, as a first step we simulate two FitzHugh-Nagumo neurons that are mutually coupled, with a periodic sub-threshold signal applied to one of them. Despite lacking a realistic biophysical simulation of neuronal coupling, model simulations yield theoretical insights that suggest that the neuron that perceives the signal can still encode the information, as it fires a spike train which has more and less expressed spike patterns whose probabilities still depend on the signal's features. 

\section*{Results}

We simulate two coupled FHN neurons as described in \textit{Methods}, with a periodic subthreshold signal that is applied to one of the neurons, referred to as neuron 1. 

\begin{figure}[!ht]
\centering
\includegraphics[width=0.7\columnwidth]{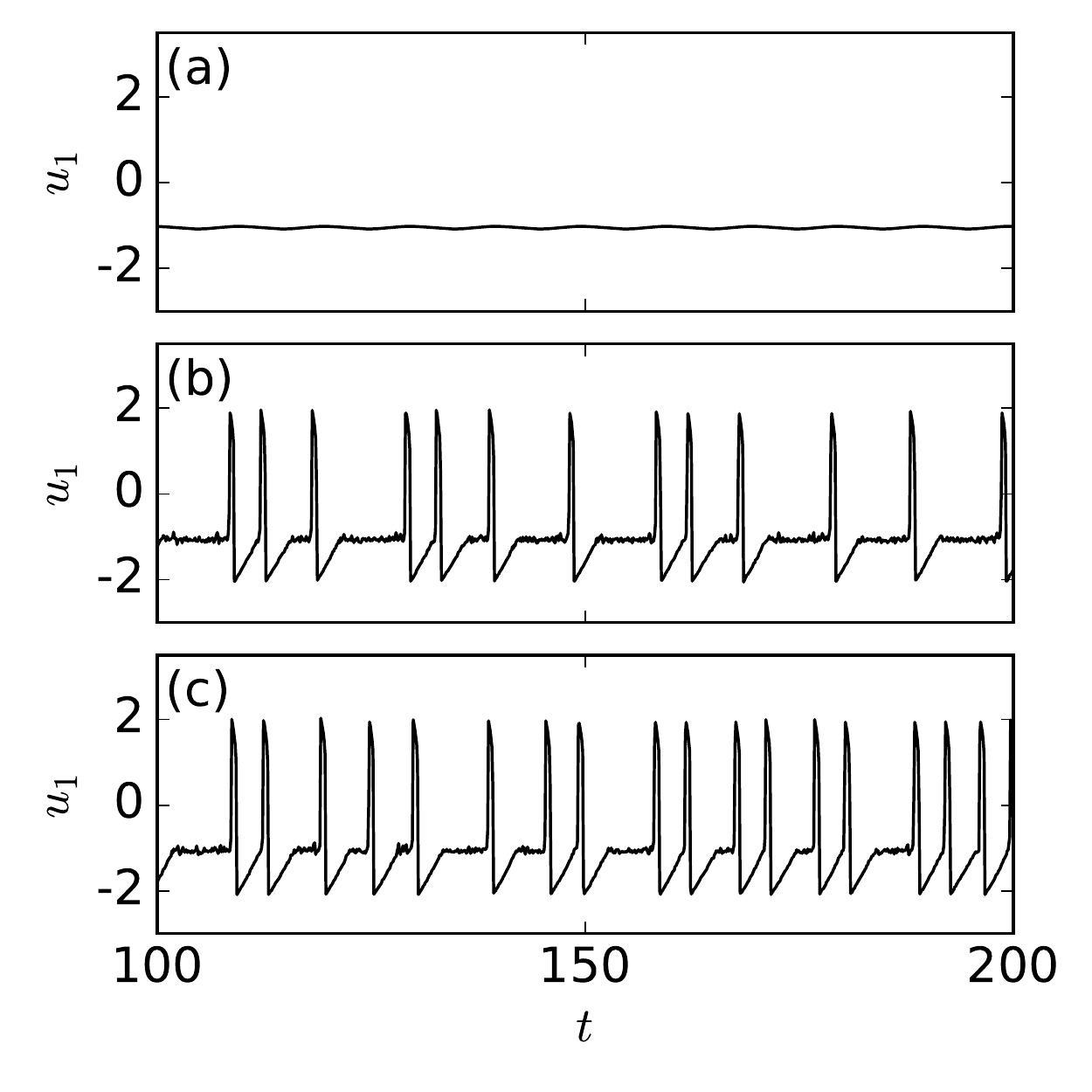}
\caption{{\bf Time-series of the voltage-like variable of neuron 1} when (a) the signal is applied, and there is no noise and no coupling (subthreshold oscillations are observed); (b) when the signal is applied and there is noise but no coupling (noise-induced spikes are observed, which carry information about the applied subthreshold signal) and (c) when the signal is applied and there is noise and coupling (an increase of the spike rate is observed).
The parameters are $a_0 = 0.05$, $T = 10$ and (a) $D = 0$, $\sigma_2 = 0$; (b) $D = 2\cdot 10^{-6}$, $\sigma = 0$; (c) $D = 2\cdot 10^{-6}$, $\sigma = 0.05$.}
\label{Fig1}
\end{figure}

Figure~\ref{Fig1} displays the voltage-like variable of neuron 1, $u_1$, in different situations. When there is no noise, no signal and no coupling, the neuron is in the rest state and when the sub-threshold signal is applied, $u_1$ displays small oscillations [panel (a)]; when noise is added, the neuron fires a spike train [panel (b)]; when the coupling to neuron 2 is added, a noticeable effect is the increase of the firing rate [panel (c)]. The differences that are qualitatively observed in these time-series are going to be quantitatively addressed by using the methods of analysis presented in \textit{Methods}. 
 
\begin{figure}[!ht]
\centering
\includegraphics[width=0.8\columnwidth]{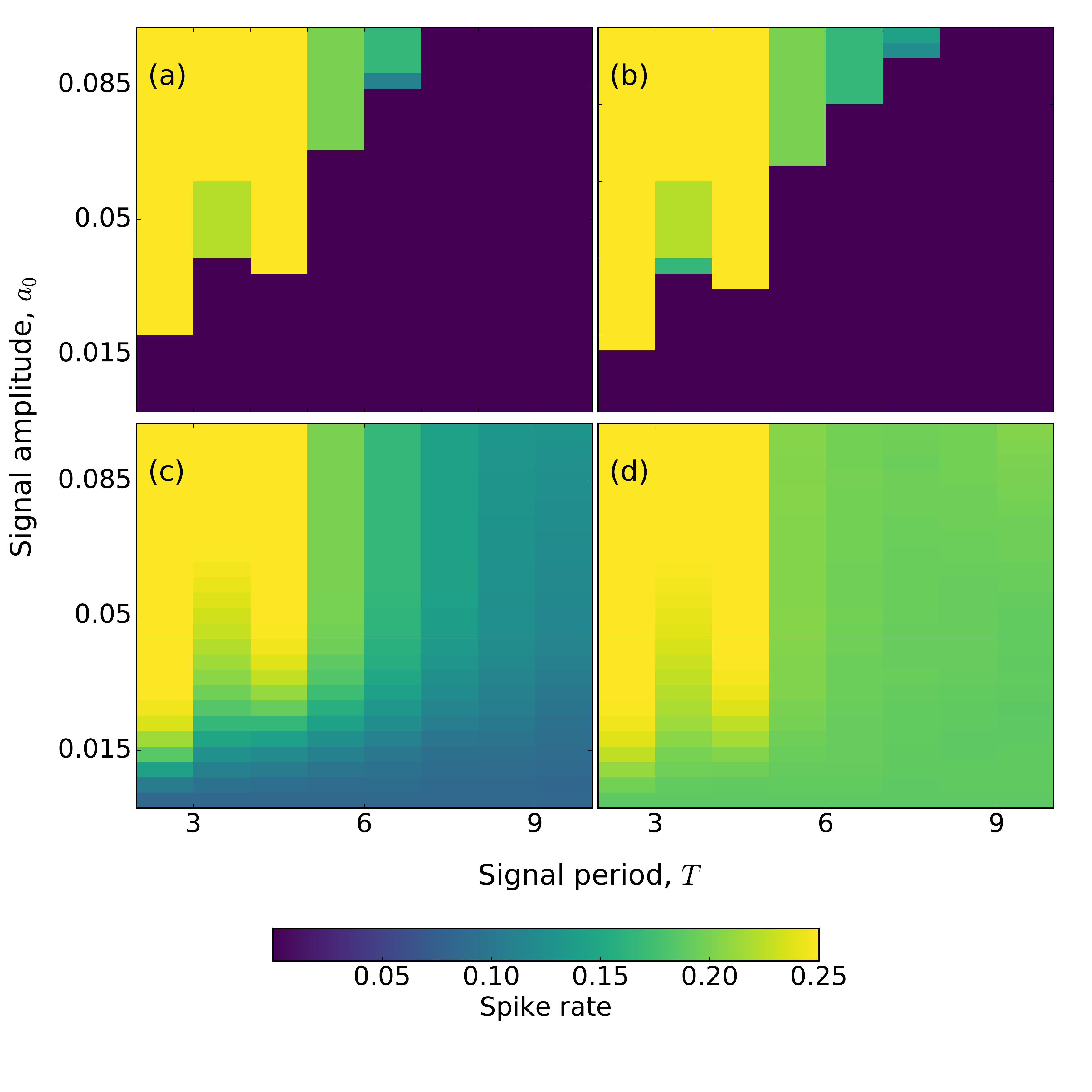}
\caption{{\bf Influence of the signal parameters on the spike rate.} The spike rate of neuron 1 in color code is plotted as a function of the signal amplitude, $a_0$, and period, $T$. Panels (a) and (b) display the deterministic spike rate ($D=0$) without coupling ($\sigma_1=\sigma_2=0$) and with coupling ($\sigma_1=\sigma_2=0.05$), respectively. In panels (c) and (d) noise is included ($D = 2\cdot 10 ^{-6}$) and it is observed that coupling (panel d) increases the spike rate with respect to the uncoupled noisy neuron (panel c).}
\label{Fig2}
\end{figure}

As we are interested in the encoding of weak signals, we first have to distinguish between a sub-threshold and a super-threshold signal. The first one refers to a signal which, in the absence of noise, it does not induce any spike [$u_1$ displays small oscillations, as in Fig.~\ref{Fig1}(a)], while the second one is a signal that is strong enough to induce spikes. A periodic signal can be either sub-threshold or super-threshold depending on both, the period and the amplitude. Thus, to identify the parameters where the signal is sub-threshold, in Fig.~\ref{Fig2} we plot the spike rate (i.e., $1/\langle I\rangle$, in color code), as a function of $a_0$ and $T$. In panel (a) neuron 1 is isolated ($\sigma_2=0$), while in panel (b) it is coupled to neuron 2 ($\sigma_1=\sigma_2=0.05$). 

When the neuron is uncoupled, for large amplitude and/or small period the signal is super-threshold, otherwise is sub-threshold. When the neuron is coupled to neuron 2 (here we want to remark that neuron 2 does not see the signal), we note that the super-threshold region is slightly larger in the parameter space ($a_0$, $T$), as compared to the uncoupled case. 

When we include noise, Figs.~\ref{Fig2}(c) and (d), we first note that in the super-threshold region (yellow) the spike rate does not change significantly (it is about the same as in panel (a), where $D$=0 and $\sigma_1=0$). This is due to the fact that in this region the spikes are  induced by the signal, while the noise or the coupling do not have a significant effect. 

In contrast, in the sub-threshold region, comparing the uncoupled (panel c) and the coupled (panel d) situations, we note that coupling significantly increases the spike rate (it almost doubles). Therefore, in this region coupling plays the role of an extra source of noise (as in this region, both, noise and coupling induce spikes).

\begin{figure}[!ht]
\centering
\includegraphics[width=0.8\columnwidth]{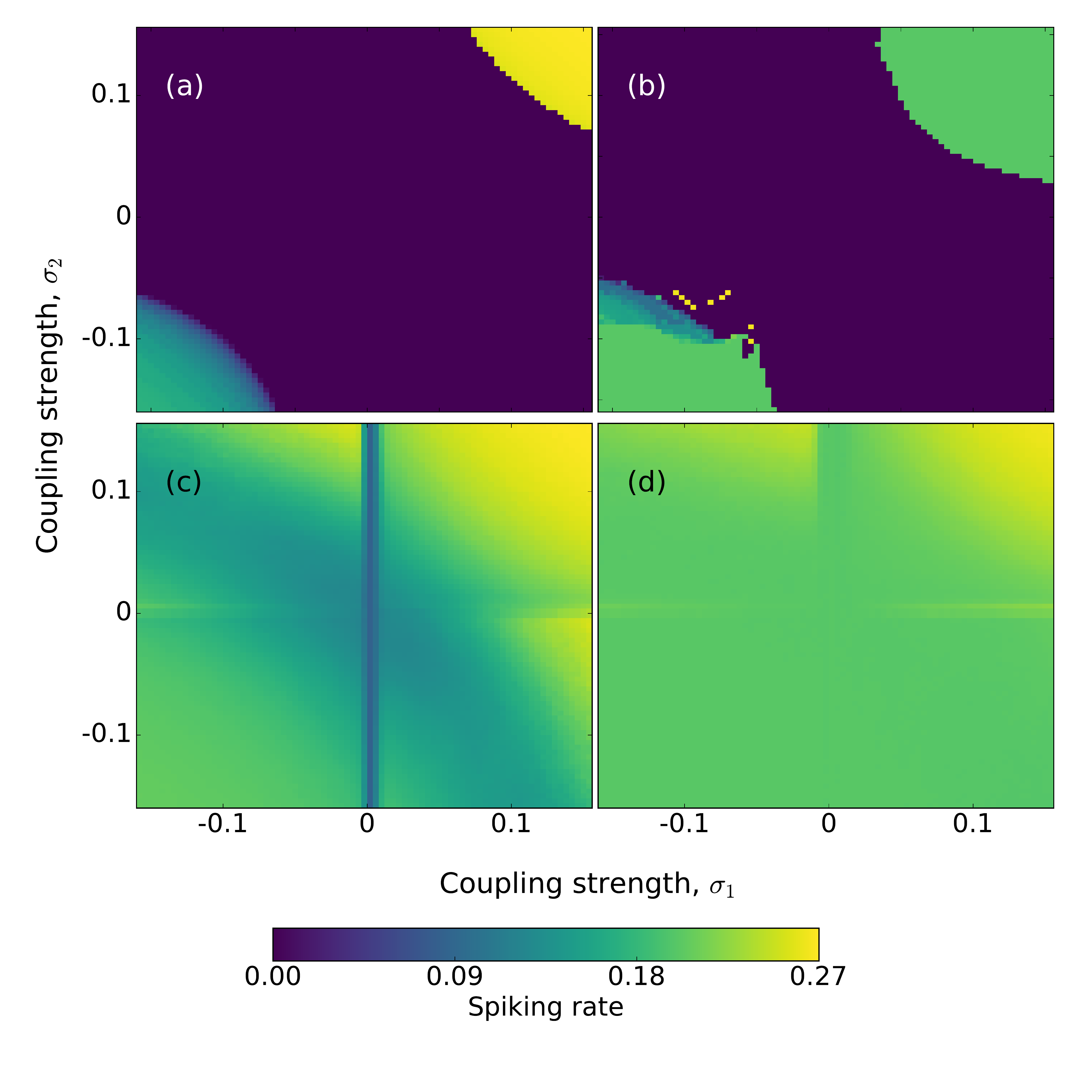}
\caption{{\bf Influence of the coupling strengths in the spike rate.} The spike rate of neuron 1 in color code is plotted as a function of $\sigma_1$ and $\sigma_2$, with and without noise: panels (a) and (b) display the deterministic spike rate ($D = 0$), while panels (c) and (d) display the spike rate of noisy neurons ($D = 2\cdot 10 ^{-6}$). In (a), (c) the signal is not applied ($a_0 = 0$) while in (b), (d) it is applied ($a_0 = 0.05$ and $T = 10$). We note that, without noise, strong enough coupling induces spikes. This occurs when $\sigma_1$ and $\sigma_2$ are both positive or both negative, regardless of the input signal. When there is noise, the effect is still present when there is no signal (in panel c the spike rate is higher when $\sigma_1,\sigma_2>0$ or when $\sigma_1,\sigma_2<0$) while it is almost washed out when the signal is applied (in panel d). The vertical line in panel (c) is due to the fact that when $\sigma_1=0$, neuron 1 is uncoupled from neuron 2, therefore, its firing rate does not depend on $\sigma_2$, which is the strength of $1\rightarrow 2$ coupling.}
\label{Fig3}
\end{figure}

Having identified the sub-threshold region in the parameter space ($a_0$, $T$) when the coupling coefficients are kept fixed ($\sigma_1=\sigma_2=0.05$), we next turn our attention to the influence of the coupling coefficients, now keeping the signal parameters fixed: we choose $a_0 = 0.05$ and $T = 10$, which are within the sub-threshold region in Fig.~\ref{Fig2}(a). Figure~\ref{Fig3} displays the spike rate as a function of $\sigma_1$ and $\sigma_2$ in different situations. In panel (a) there is no signal and no noise. We observe that when both $|\sigma_1|$ and $|\sigma_2|$ are large enough, the coupling induces spikes. Thus, a sub-threshold region in the parameter space $(\sigma_1, \sigma_2)$ is observed. Positive coupling coefficients result in higher spike rate, in comparison with negative coefficients. In panel (b), the noise is still zero but the signal is applied. Here we note that the size of the super-threshold region is slightly larger in comparison to panel (a), and now positive and negative coupling coefficients produce similar spike rates. Figures~\ref{Fig3} (c) and (d) display the spike rate when noise is included, without and with signal respectively. The vertical line in panel (c) is due to the fact that when $\sigma_1=0$ neuron 1 is uncoupled from neuron 2, and thus its spike rate does not depend of $\sigma_2$. Without signal, positive coupling coefficients result in larger spike rate as compared to negative ones, however, when the signal is applied these differences are washed out. 

In the following and unless otherwise stated, in order to limit the number of parameters we take $\sigma_1 = \sigma_2 = \sigma$. As well, we will use $\sigma=0.05$, $a_0 = 0.05$ and $T = 10$. For these parameters the signal and the coupling act as sub-threshold perturbations: without noise neuron 1 does not fire any spike.

\begin{center}
\begin{figure}[!ht]
\includegraphics[width=\columnwidth]{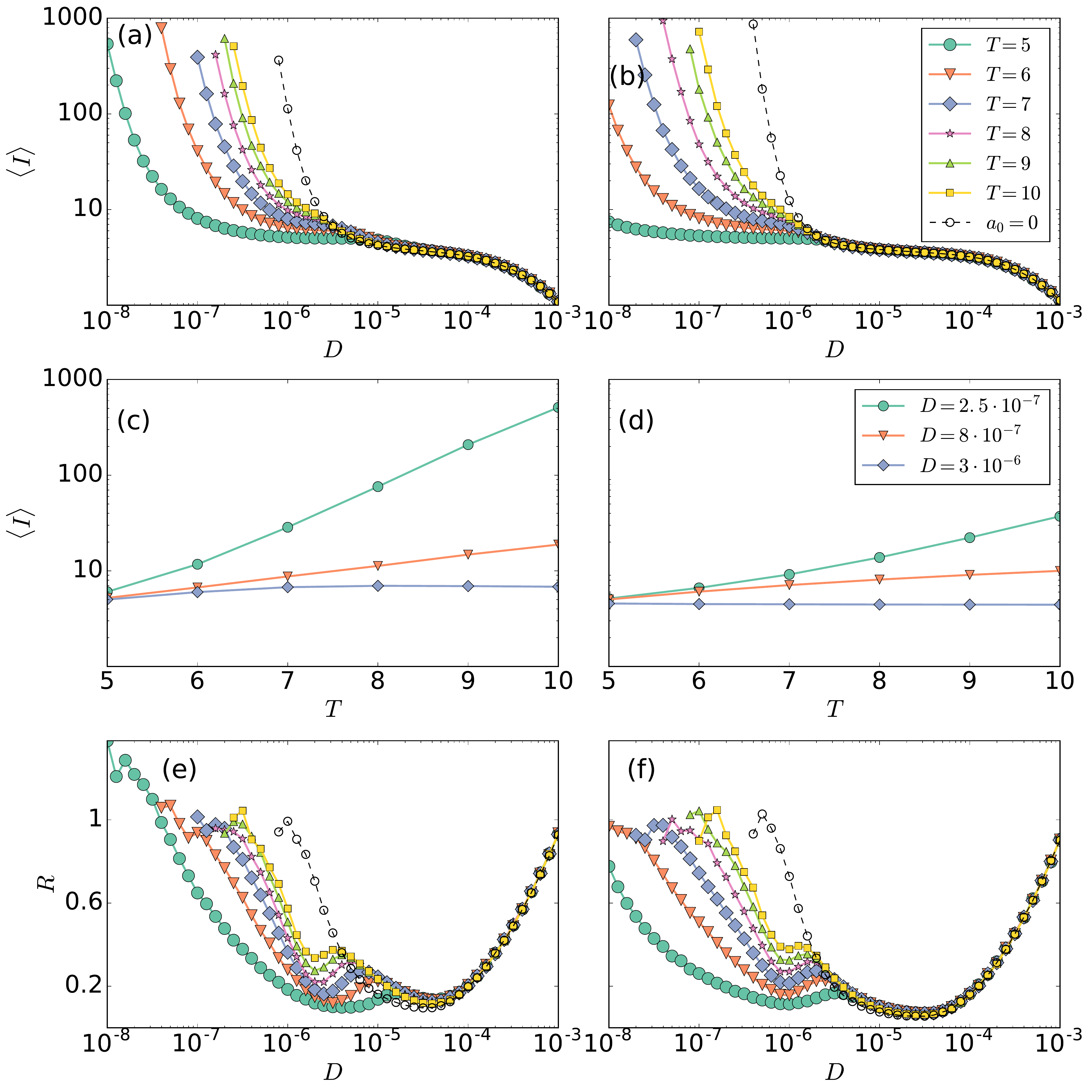}
\caption{{\bf Interplay of noise and the period of the signal.} (a), (b) Mean inter-spike interval, $\left<I \right>$, of neuron 1 as a function of the noise strength, for different periods of the external signal; (c), (d) $\left<I \right>$ vs. the period of the signal and (e), (f) Normalized standard deviation of the ISI distribution, $R$, as a function of the noise strength, for different periods of the signal. Panels (a), (c) and (e) are without coupling ($\sigma_1 =\sigma_2 = 0$), while (b), (d) and (f) are with coupling ($\sigma_1 =\sigma_2 = 0.05$). In panels (a) and (b) we note that, for strong enough noise, the mean ISI does not depend of the period of the signal. In panels (c), (d) we note that for weak and moderate noise, $\left<I \right>$ increases linearly with $T$, while for strong noise, $\left<I \right>$ saturates to the refractory period, $T_e$ (i.e., the duration of the excursion in the phase space when a large enough perturbation triggers a spike), which is nearly independent of $T$. In panels (e), (f) we see two minima, one that occurs when $\left<I \right>\sim T$, which is interpreted as due to stochastic resonance \cite{sr,sr_chialvo_longtin_1997,chialvo_longtin_1998}, and another that occurs when $\left<I \right>\sim T_e$, which is interpreted as due to coherence resonance \cite{PIK97,not_coh_res}}.
\label{Fig4}
\end{figure}
\end{center}

To further characterize the role of noise, Fig.~\ref{Fig4} displays the mean inter-spike interval, $\langle I \rangle$, as a function of noise intensity for different periods of the applied signal. In panel (a) $\sigma = 0$, while in panel (b), $\sigma = 0.05$. For both cases there is clearly a noise dominated regime, where $\langle I \rangle$ is the same, regardless of the coupling and of the period of the signal. In contrast, for low noise levels the coupling and the period affect the $\langle I \rangle$. 

Regarding the role of the period of signal, when the noise level is low, the larger $T$ is, the larger $\langle I \rangle$ is. There is a linear relation, as shown in Figs.~\ref{Fig4}(c) and (d), which holds for both, the coupled and the uncoupled cases. For stronger noise, $\langle I \rangle$ remains constant when increasing $T$. 

In panel (a) ($\sigma = 0$) we can also compare the mean ISI when the signal is applied (solid symbols indicate $a_0\ne0$ and different periods) and when the signal is not applied (empty circles): we see that, when $a_0\ne0$ the neuron starts firing at lower noise intensities as compared to $a_0=0$. Comparing panel (a) with panel (b) ($\sigma = 0.05$) we note that when neuron 1 is coupled to neuron 2, it starts firing at even lower noise intensities. 

Noise-induced regularity in the spike train \cite{PIK97,jgo_phys_rep_2004,noise_in_neural} is characterized in panels (e) and (f), where the normalized standard deviation of the ISI distribution, $R$, is plotted against the noise intensity for different $T$, without and with coupling, respectively. In both panels, two minimums are observed. Whereas the first one indicates stochastic resonance \cite{sr,sr_chialvo_longtin_1997,chialvo_longtin_1998}, as it occurs when $T \sim \langle I \rangle$, the second one reveals the coherence resonance phenomenon~\cite{PIK97,not_coh_res}, which is independent from the period of the signal. It occurs for an intermediate value of the noise amplitude for which noise-induced oscillations become most coherent. For some periods $T$ a maximum appears for very small values of the intensity of the noise. Such maxima are a signature of anticoherence resonance \cite{Lacasta02}. 

After having characterized the role of the various parameters in the spike rate, we next apply non-linear ordinal analysis in order to undercover possible preferred spike patterns. 

\begin{figure}[!ht]
\center
\includegraphics[width=0.9\columnwidth]{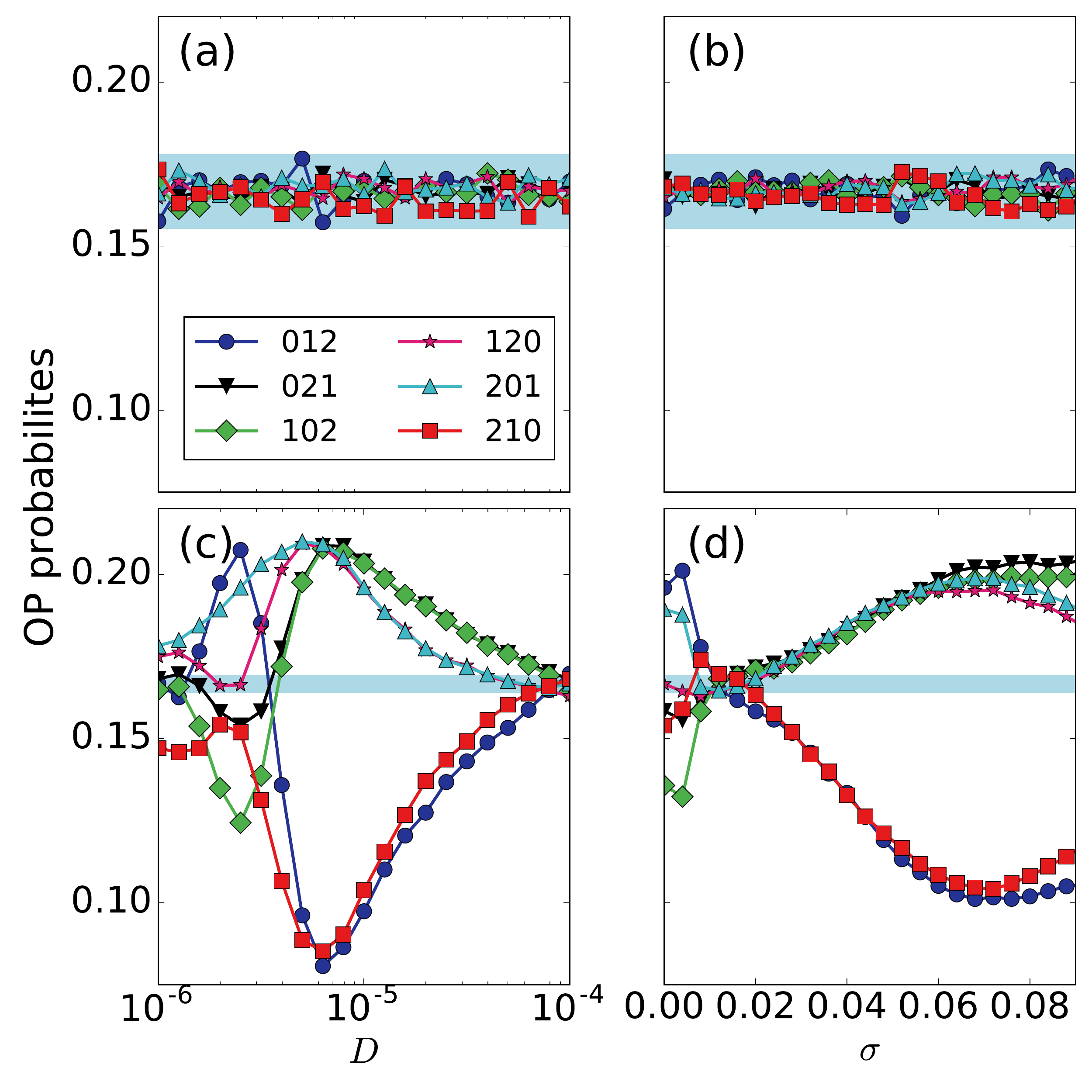}
\caption{{\bf Ordinal probabilities as a function of the noise and coupling strengths.} In panels (a), (b) the probabilities of the six ordinal patterns are plotted respectively as a function of $D$ (for $\sigma_1=\sigma_2=0$) and as function of $\sigma$ (for $D = 2\cdot 10^{-6}$), both for $a_0 = 0$. Panels (c) and (d) are as (a), (b), but a subthreshold signal is applied ($a_0 = 0.05$ and $T = 10$). In all the panels the gray region indicates the interval of probability values that are consistent with the uniform distribution with 99.74\% confidence level. We observe that without the signal [panels (a) and (b)], there are no noise-induced or coupling-induced ISI correlations, as all the ordinal probabilities are within the gray interval of values. In contrast, when the signal is applied [panels (c) and (d)], the probabilities are not consistent with the uniform distribution. In these panels we also note that the variation of the ordinal probabilities with $D$ or with $\sigma$ is qualitatively similar. This similarity is valid for low $D$ or low $\sigma$ values.}
\label{Fig5}
\end{figure}

We begin by considering the situation in which no signal is applied and analyze the effect of increasing the noise level or the coupling strength: Figs.~\ref{Fig5} (a) and (b) display the ordinal probabilities as a function of $D$ and $\sigma$, respectively. We note that neither the noise nor the coupling induce temporal correlations along the ISI sequence (as all the probabilities are within the gray region that indicates values consistent with equal probabilities). When the signal is applied, panels (c) and (d), we note that increasing either the noise level or the coupling strength induce temporal ordering in the ISI sequence, as the probabilities and not consistent with the uniform distribution and thus reveal the presence of preferred and less frequent spike patterns. Moreover, we note that the variation of the probabilities with $D$ or $\sigma$ is qualitatively similar. 

\begin{figure}[!ht]
\center
\includegraphics[width=0.9\columnwidth]{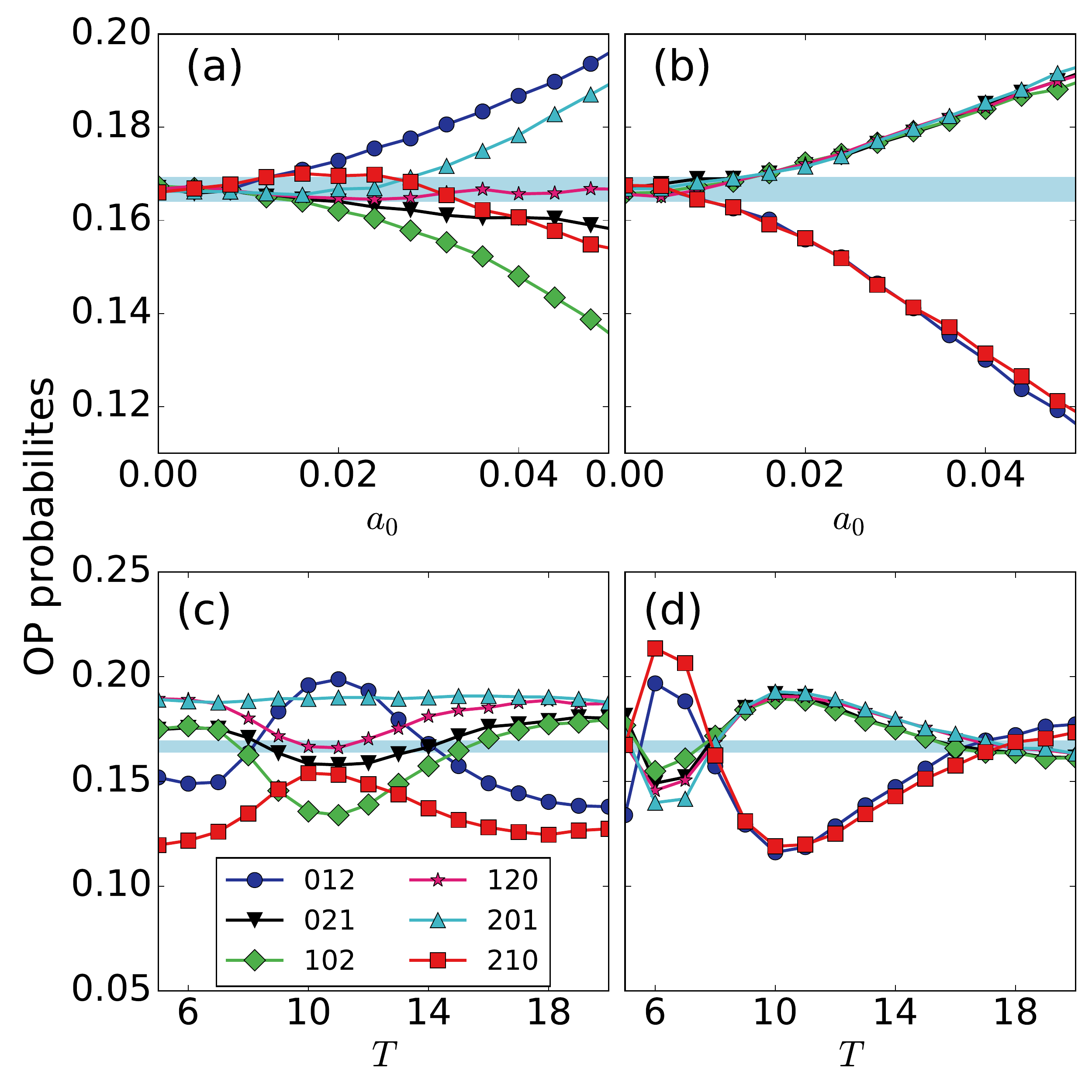}
\caption{{\bf Influence of coupling on signal encoding.} Panels (a) and (b) display the ordinal probabilities as a function of $a_0$ without and with coupling, respectively. Panels (c) and (d) display the probabilities as a function of $T$ without and with coupling, respectively.  In (a) and (b) $T = 10$, in (c) and (d) $a_0 = 0.05$. In all panels the noise strength is $D = 2\cdot 10^{-6}$. In (a) and (c) $\sigma = 0$, in (b) and (d) $\sigma = 0.05$. Comparing panels (a) and (b) we note that, with coupling, the ordinal probabilities are outside the blue region (that indicates the interval of values that are consistent with the uniform distribution with 99.74\% confidence level) for lower values of $a_0$. This means that, when neuron 1 is coupled to neuron 2, is able to detect and encode signals with smaller amplitude.   Comparing panels (c) and (d) we note that, with or without coupling, the probabilities depend of the period of the signal. This suggests that the encoding mechanism is robust to coupling, as the neuron that perceives the signal can still encode the information about the period, by firing a spike sequence which has more frequent and less frequent patterns, which depend on the signal period.}
\label{Fig6}
\end{figure}

Next, we investigate how the coupling affects the encoding of the signal features (the amplitude and period): we compare how the ordinal probabilities vary with $a_0$ and $T$, when neuron 1 is isolated [Figs. \ref{Fig6} (a) and (c)] and when it is coupled to neuron 2 [Figs. \ref{Fig6} (b) and (d)]. In both cases, when $a_0$ increases (within the subthreshold region) the probabilities monotonically increase or decrease. This variation is consistent with the results reported in \cite{REI16}. It is important to remark that in \cite{REI16} the sub-threshold signal was applied to the slow variable, $v$, while here it is applied to the fast variable, $u$. In both cases, the probabilities encode information of the amplitude of the signal. Nevertheless, coupling to neuron 2 changes the preferred and infrequent patters, i.e., modifies the temporal order in the spike sequence. For instance, for $\sigma = 0.05$ the probability of the ordinal pattern 012 monotonically increases with $a_0$, whereas for $\sigma = 0.05$ monotonically decreases. In panels (b) and (d) we note that, with or without coupling, the preferred and infrequent patterns depend on the period of the signal, confirming the results reported in \cite{REI16}.  

\begin{figure}[!ht]
\center
\includegraphics[width=\columnwidth]{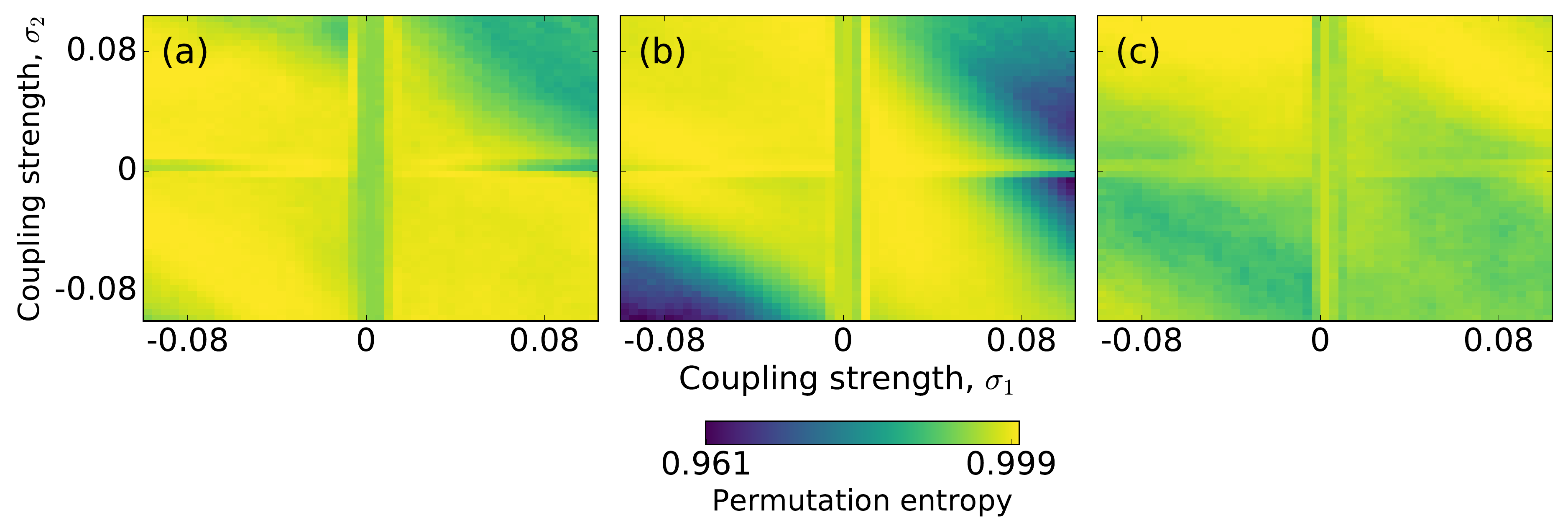}
\caption{{\bf Influence of the coupling strengths on signal encoding.} The information content of the sequence of ordinal patterns computed from the spikes of neuron 1 is quantified by the permutation entropy in color code that is plotted as a function of the coupling strengths $\sigma_1$ and $\sigma_2$ for three periods of the signal: $T = 6$, $T = 10$ and $T = 14$, panel (a), (b) and (c), respectively. Other parameters: $a_0 = 0.05$ and $D = 2 \cdot 10^{-6}$. We note that the information content is maximum (lower entropy) for an intermediate value of $T$ and coupling strengths such that  $\sigma_1\sigma_2 > 0$.}
\label{Fig7}
\end{figure}

Next we address the issue whether there is an optimal coupling configuration (i.e., a set of coupling coefficients $\sigma_1$ and $\sigma_2$) for signal encoding. To quantify the information content of the spike train, when is represented by symbolic ordinal patterns constructed from ISI intervals, we calculate the entropy computed from the probabilities of the ordinal patterns (known as \textit{permutation entropy}, $H = - \sum_i p_i \log{p_i}$ \cite{BAN02}). To investigate how the coupling coefficients that maximize the information content (i.e. minimize the entropy) depend on the input signal, we calculate the entropy for different periods. Fig. \ref{Fig7} displays the permutation entropy (normalized to its maximum value) in color code as a function of $\sigma_1$ and $\sigma_2$ for three periods: $T = 6$, $T = 10$ and $T = 14$, panels (a), (b) and (c), respectively.  We observe that for small and large periods ($T = 6$ and $T = 14$) and for all coupling strengths, the entropy is close to 1, which indicates that the ordinal probabilities are all similar, i.e., neuron 1 has an stochastic dynamics. Whereas for $T = 10$ there is a region of coupling strengths where lower entropy values reveal that there are more likely and less likely patterns, i.e., the spike sequence carries information about the signal. From panel (b) we learn than when $\sigma_1\sigma_2 > 0$ the coupling to a second neuron helps to encode the signal, as the entropy has lower values. In contrasts, when $\sigma_1\sigma_2 < 0$ the coupling to the second neuron detriments the encoding of the signal, because the permutation entropy is highest. 

\begin{figure}[!ht]
\center
\includegraphics[width=\columnwidth]{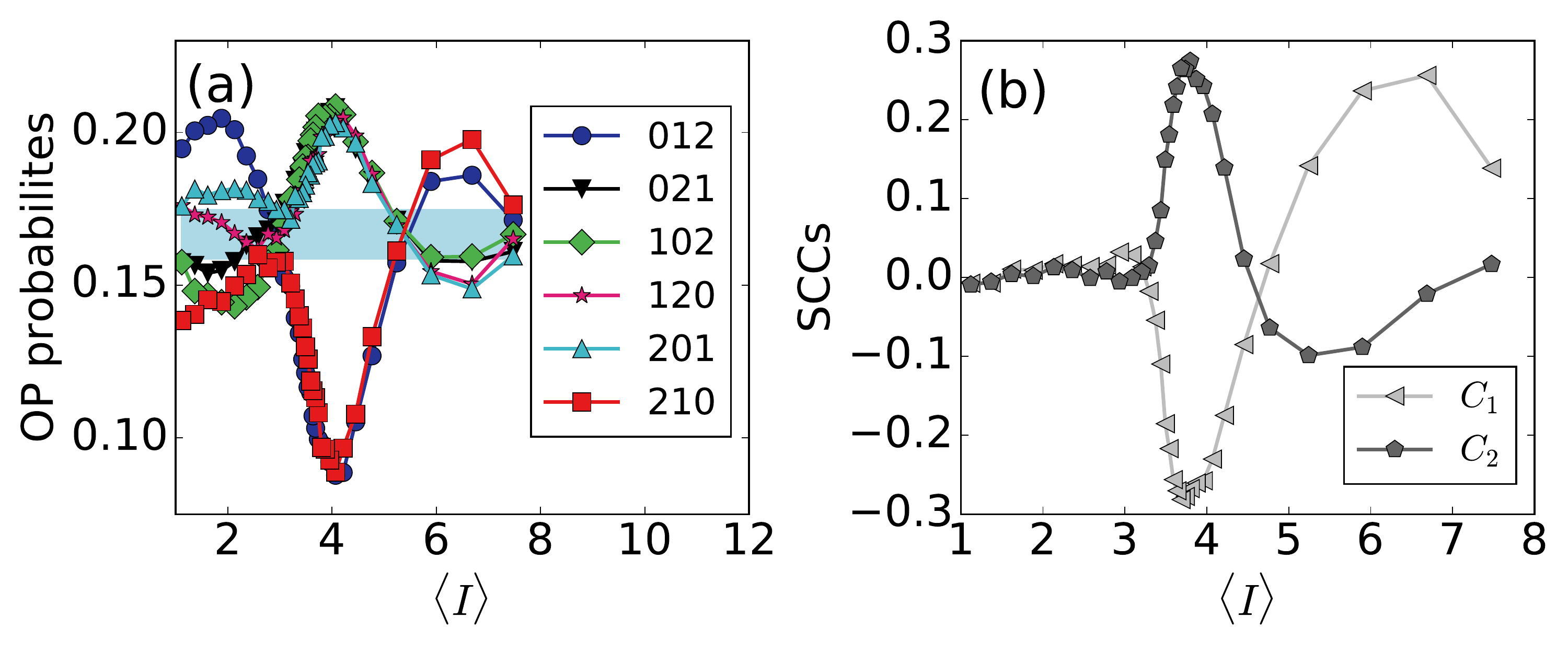}
\caption{{\bf Relation between ordinal probabilities, serial correlation coefficients and mean ISI.} (a) Ordinal probabilities and (b) serial correlation coefficients, $C_1$ and $C_2$, as a function of the mean ISI, $\langle I \rangle$, when the noise strength is varied within the range $10^{-6}  \leqslant D \leqslant 10^{-3}$. The signal parameters are $T = 8$, $a_0 = 0.05$ and the coupling strength is $\sigma = 0.05$.}
\label{Fig8}
\end{figure}

Classical measures to quantify linear ISI correlations are the serial correlation coefficients (SCCs, see \textit{Methods}). Next, we compare the results obtained with nonlinear symbolic ordinal analysis, with those obtained with SCCs. To do this, we first compare in Fig.~\ref{Fig8} how the ordinal probabilities and the SCCs vary while changing the mean ISI  (we calculated the mean ISI $\langle I \rangle$ for each noise intensity within the range $10^{-6}  \leqslant D \leqslant 10^{-3}$) for a fixed period $T$. We see that while the probabilities of ordinal patterns 012 and 210 (respectively three increasingly and decreasingly spikes) show a minimum at $\langle I \rangle = 4$ the other four show a maximum. This is captured as well with the linear measures $C_1$ and $C_2$, which respectively show a minimum and a maximum at $\langle I \rangle \approx 4$. Nevertheless correlations that appear for large noise, i.e. small $\langle I \rangle$,  which  are captured by ordinal patterns probabilities (they are outside the blue region) are not captured by the linear measures $C_1$ and $C_2$.

\begin{figure}[!ht]
\center
\includegraphics[width=0.8\columnwidth]{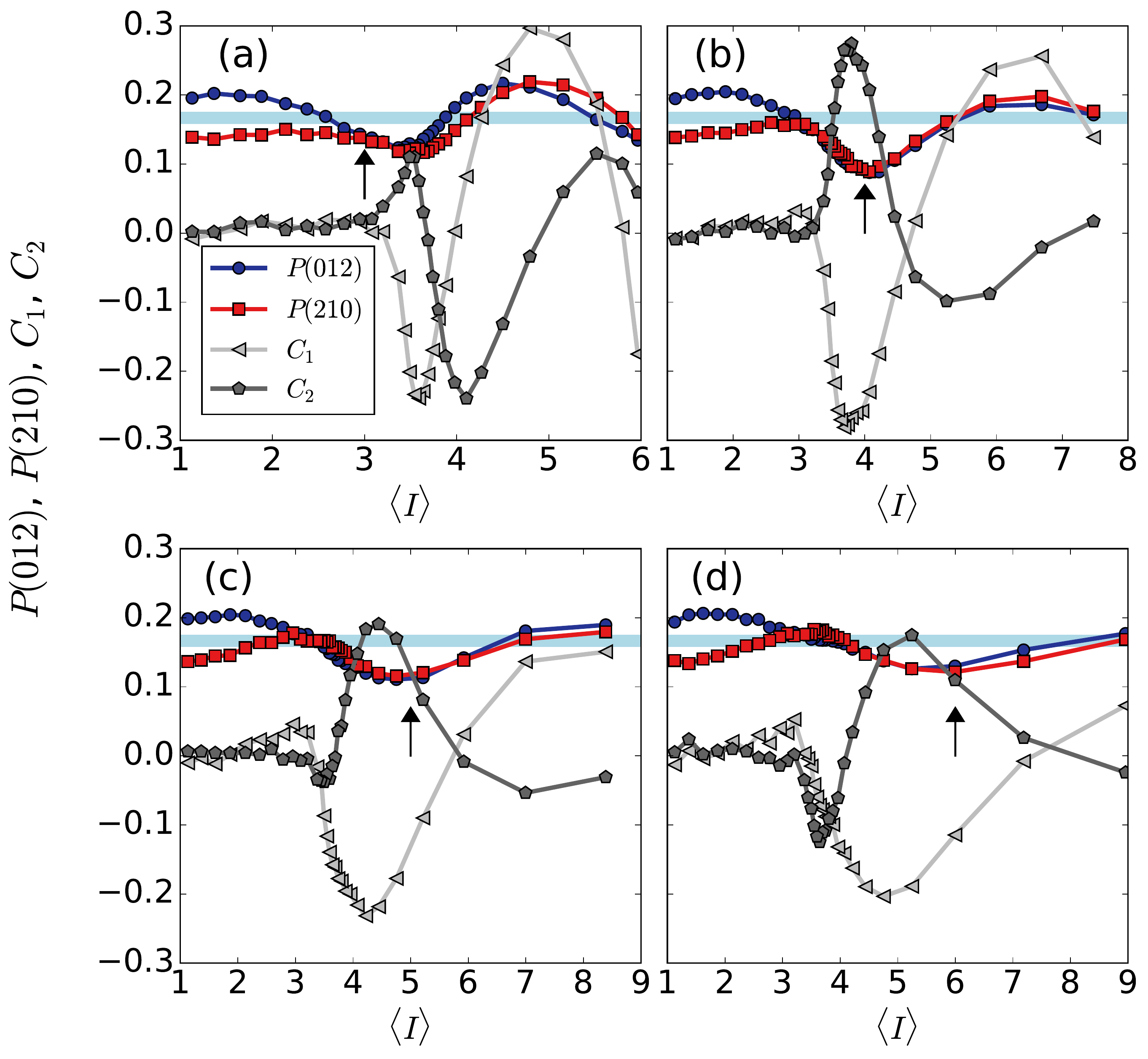}
\caption{{\bf} {\bf Comparison of results of linear and nonlinear measures.} $P(012)$, $C_1$ and $C_2$ in function of $\langle I \rangle$ for different periods $T = 6$ , $T = 8$, $T = 10$  and $T = 12$ in panels (a), (b), (c) and (d), respectively. Noise amplitude was within the range $10^{-6}  \leqslant D \leqslant 10^{-3}$, $a_0 = 0.05$ and $\sigma = 0.05$.}
\label{Fig9}
\end{figure}

Next, we choose the trend patterns 012 and 210 (three increasingly longer or shorter ISIs), and analyze how their probabilities vary with the mean ISI, and compare with the variation of $C_1$ and $C_2$. Our goal is, first, to determine if there is any relation between the linear quantifiers of ISI correlations, $C_1$ and $C_2$, and the nonlinear ones, $P(012)$ and $P(210)$. Secondly, we want to analyze how they depend on $\langle I \rangle$ and $T$. Figure~\ref{Fig9} displays $P(012)$, $P(210)$, C$_1$ and $C_2$ as a function of  $\langle I \rangle$ for four different periods $T = 6, 8, 10$ and $12$ in panels (a), (b), (c) and (d), respectively.  A first thing we note is that the minimum of $P(012)$ and $P(210)$ tends to occur when $T \sim \langle I \rangle/2$ (black arrows in the different panels indicate $T = \left<I \right>/2$). We also note that when $\langle I \rangle$ is too short (i.e., the noise level is high), $C_1$ and $C_2$ are close to zero, regardless of the period of the signal; in contrast, $P(012)$ and $P(210)$ are not within the region of values which are consistent with uniform probabilities, and thus, carry information about the subthreshold signal.

\begin{figure}[!ht]
\center
\includegraphics[width=0.7\columnwidth]{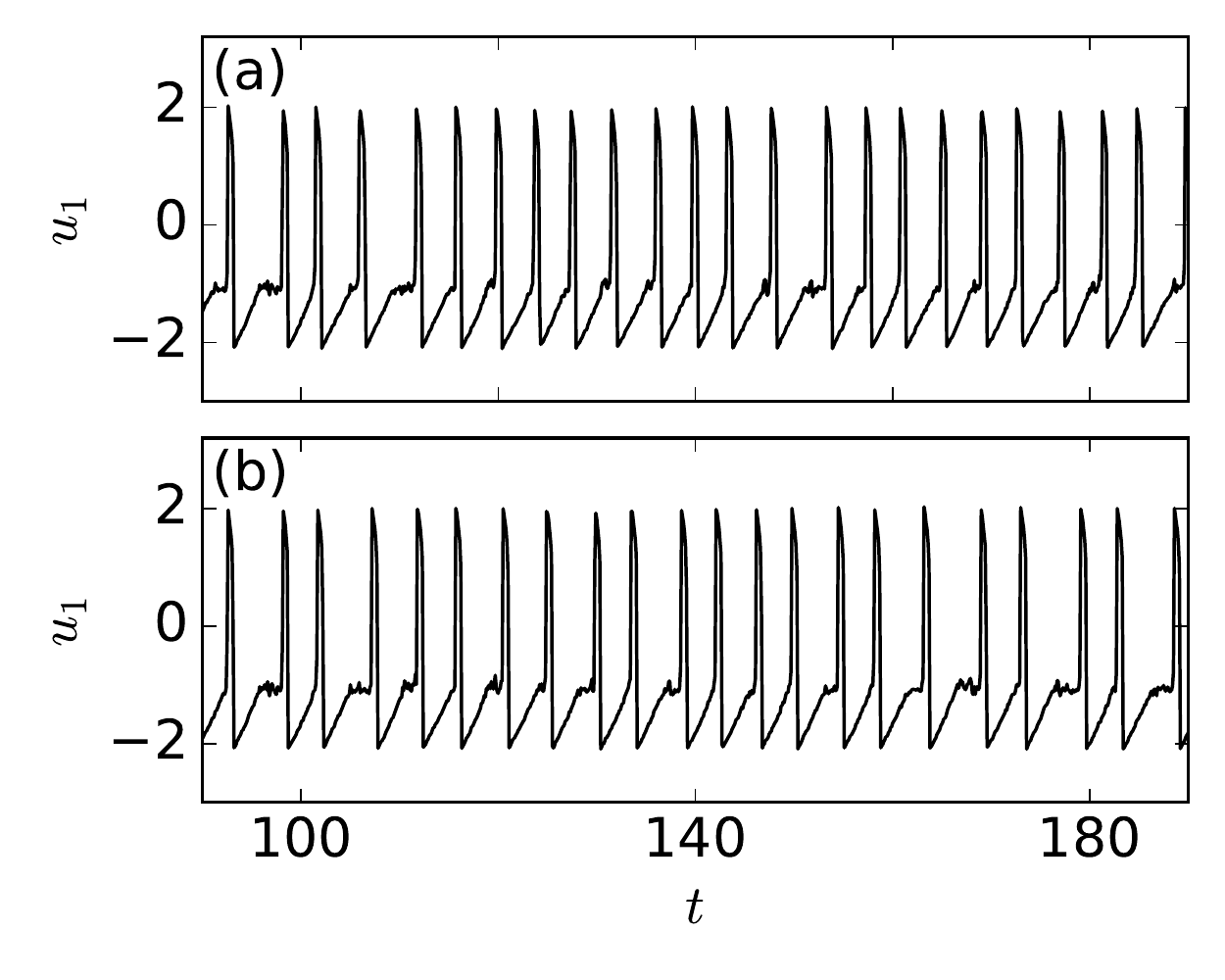}
\caption{{\bf Examples of spike sequences where pattern 012 is more/less expressed.}
Spike train of neuron 1 when the model parameters are such that the ordinal pattern 012 (i.e., three increasingly separated spikes) is more expressed (a) ($P(012) = 0.22$) and less expressed (b) ($P(012) = 0.08$). In (a) $T = 6$ and in (b) $T = 8$. Other parameters are $\sigma = 0.05$, $a_0 = 0.05$ and $D = 3.2\cdot 10^{-6}$.}
\label{Fig10}
\end{figure}

Figure~\ref{Fig10} displays temporal series for two different values of the signal period $T = 6$ and $T = 8$ and the same noise intensity. We observe how for $T = 6$ ordinal pattern 012 is highly expressed in contrast to the period $T = 8$, for which it is less observed.  

\begin{figure}[!ht]
\center
\includegraphics[width=0.75\columnwidth]{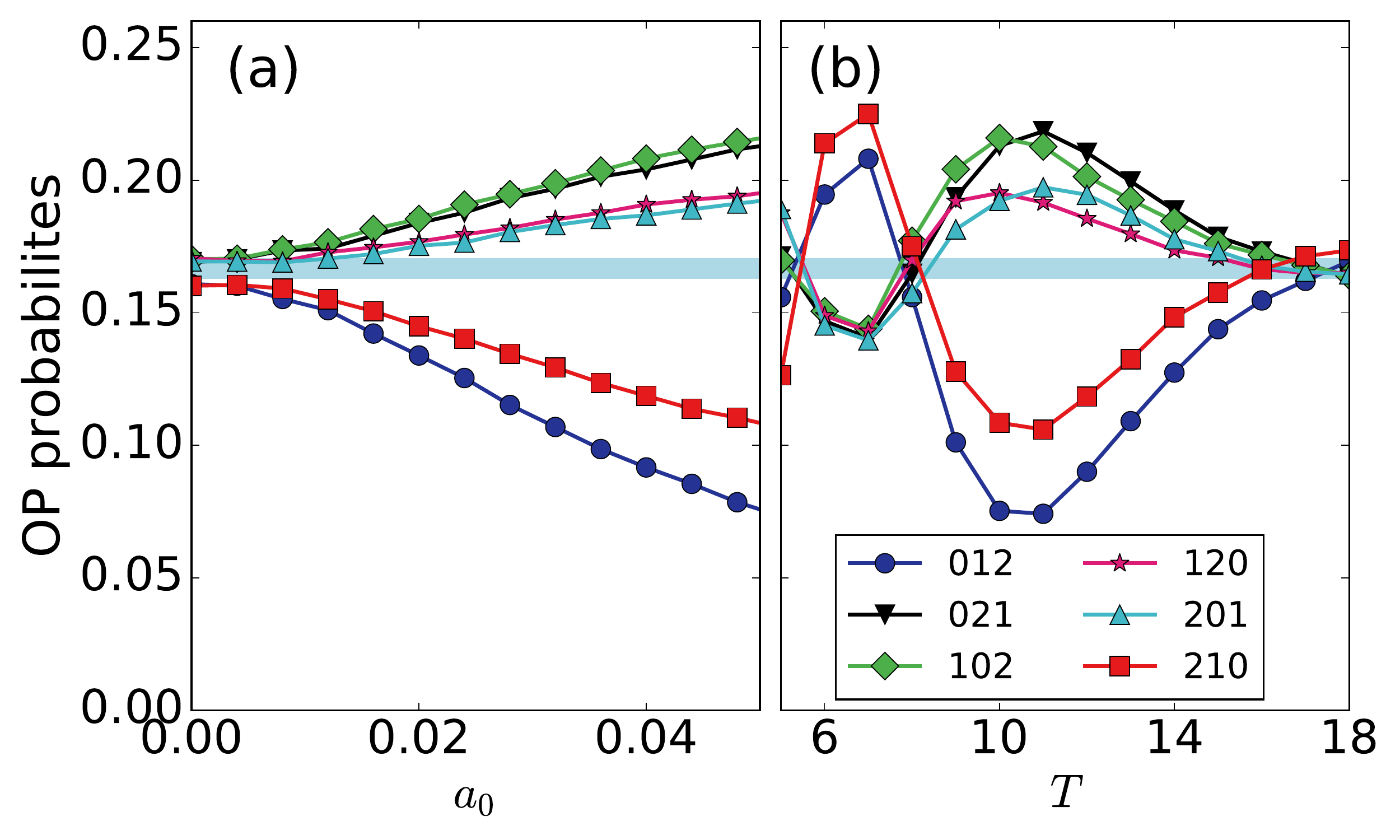}
\caption{{\bf Influence of diffusive coupling on the signal encoding.} Panels (a) and (b) display the ordinal probabilities as a function of $a_0$ (with $T = 10$) and as a function of $T$ (with $a_0 = 0.05$). Other parameters are $\sigma = 0.025$ and $D = 2\cdot 10^{-6}$. We note that the encoding of the signal features (amplitude and period) is as in Fig.~\ref{Fig6}, which was done with non diffusive coupling. }
\label{Fig11}
\end{figure}

Another relevant issue to discuss is how the coupling terms are implemented. While we have presented simulations of Eqs.~\ref{eq:model_2}, where the terms $\sigma_2 u_1$ and $\sigma_1 u_2$ couple neuron 1 to neuron 2 and vice-versa \cite{neiman_prl_2015}, we have also performed simulations with i) the coupling in the recovery-like variable (i.e., $\sigma_2 v_1$ and $\sigma_1 v_2$ added to the rate equations of $v_2$ and $v_1$ respectively) and ii) with differential coupling (i.e., $\sigma (u_1-u_2)$ and $\sigma (u_2-u_1)$ added to the rate equations of $u_1$ and $u_2$ respectively). We have consistently found that the probabilities of the ordinal patterns vary with both, the period and the amplitude of the signal, in a similar way as with with non diffusive coupling (see Fig.~\ref{Fig11}). We have also found that the relationship between $P(012)$, $P(210)$ and $\langle I \rangle$ shown in Fig.~\ref{Fig9} is robust. 

\section*{Discussion}

We have studied two coupled FitzHugh-Nagumo neurons with a subthreshold periodic signal applied to one of them. We have used symbolic analysis to investigate the spike train fired by the neuron that perceives the signal. By applying ordinal analysis to the sequence of inter-spike intervals (ISIs) we have shown that the spike train has ordinal probabilities which depend on the signal features (the amplitude and the period). By lowering the firing threshold, the second neuron facilitates the detection and encoding of the signal applied to the first neuron.  We have also shown that the ordinal probabilities achieve maximum or minimum values when the period of the external signal is about half the mean ISI. In addition, we have shown that, when the noise level is high, the ordinal probabilities encode information about the subthreshold signal, while the serial correlation coefficients (SCCs) at lag 1 and 2 vanish and mean ISI is independent of the signal period.

Our findings contribute to advance the understanding of how neurons encode information about subthreshold signals in noisy environments. The encoding mechanism demonstrated here, by which the period and the amplitude of the applied sub-threshold signal are encoded in the values of the ordinal probabilities, is very slow if the probabilities are computed from the spike train of a single neuron, because a large number of spikes are needed in order to determine the probabilities of the different spike patterns. However, if the encoding is performed by a neuronal ensemble, then, the probabilities could be computed from the spike trains of a large number of neurons, and in this case, only few spikes per neuron are be enough to compute the probabilities. This ensemble-based mechanism allows also encoding a sub-threshold signal with time-varying amplitude and/or period. Therefore, as future work, it will be interesting to extend this study to models of neuronal ensembles~\cite{brunel_2000,roxin_prl_2005,Ostojic_2014,torcini_2017}.

\section*{Materials and methods}

\subsection*{Model}

We consider two identical FitzHugh-Nagumo neurons \cite{FIT61a,NAG62}, mutually coupled as in \cite{neiman_prl_2015}, with a periodic signal applied to one of them (referred to as neuron 1): 
	\begin{equation}
	 \begin{gathered}
	\epsilon \dot{u_1}= u_1 - \frac{u_1^3}{3} - v_1  + a_0\cos(2\pi t/T) + \sigma_1 u_2 +\sqrt{2D}\xi_1(t),\\
	\dot{v_1} = u_1+ a,\\
	\epsilon \dot{u_2}= u_2 - \frac{u_2^3}{3} - v_2  + \sigma_2 u_1 +\sqrt{2D} \xi_2(t)\\
	\dot{v_2} = u_2+ a 
	\end{gathered}
	\label{eq:model_2}
	\end{equation}
	
The coupling configuration is schematically represented in Fig.~\ref{Fig11}. The dimensionless variables $u_i$ and $v_i$ are a fast variable that represents the voltage of the membrane, and a recovery-like variable that represents the refractory properties of the membrane (slow variable); $a$ and $\epsilon$ are parameters that control the spiking activity of the uncoupled neurons. The coupling terms $\sigma_2 u_1$ and $\sigma_1 u_2$ mimic synaptic currents from neuron 1 to neuron 2 and vice-versa~\cite{neiman_prl_2015}. The signal has amplitude $a_0$ and period $T$.  The noise is modeled with statistically independent Gaussian white noise terms [$\langle \xi_i(t)\xi_i(t')\rangle = \delta(t-t')$ and  $\langle \xi_i(t)\xi_j(t)\rangle = \delta(i-j)$] and the noise level, $D$, is the same for both neurons.

\begin{figure}[!ht]
\center
\includegraphics[width=0.75\columnwidth]{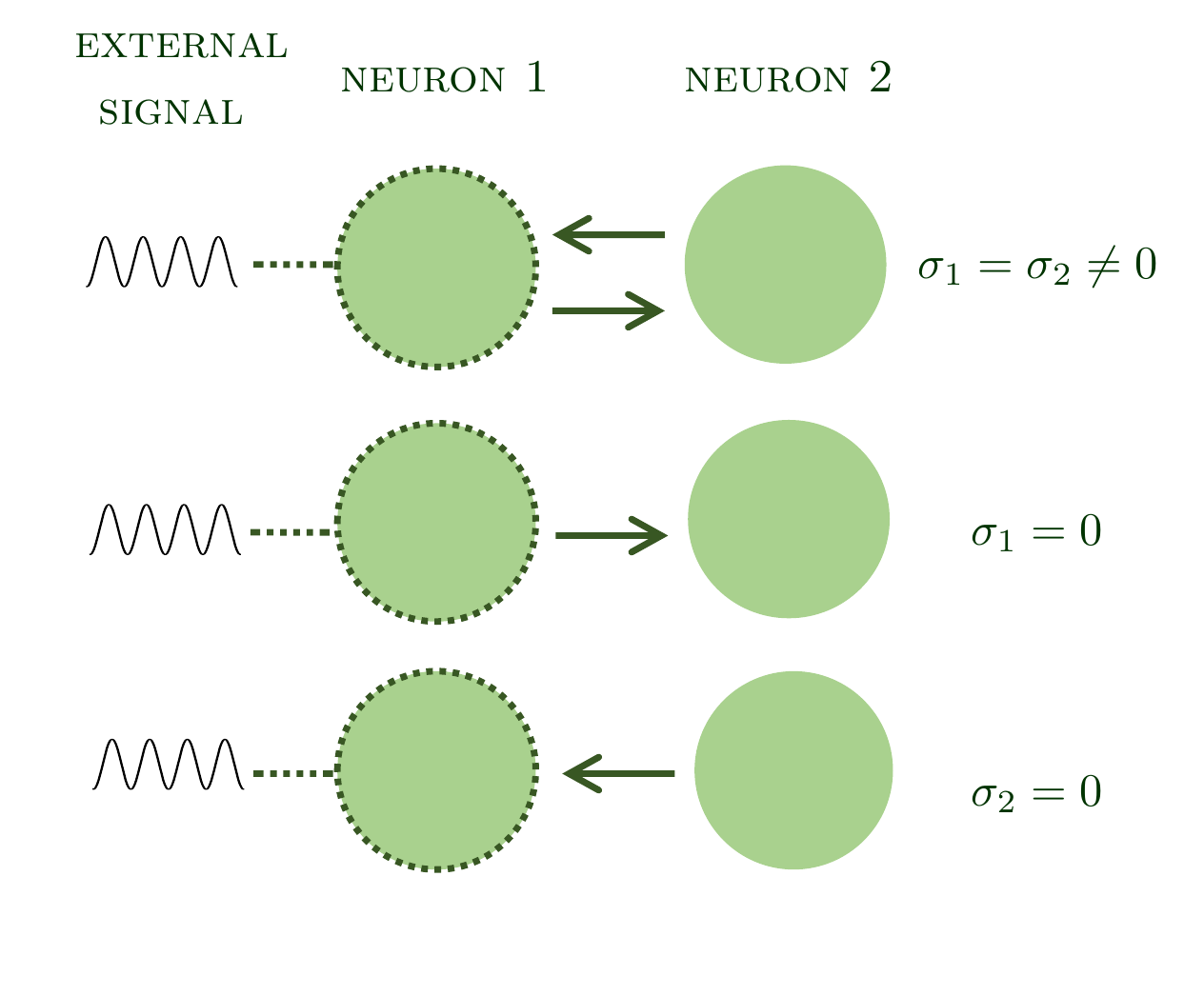}
\caption{{\bf Schematic representation of two mutually coupled neurons, one of which (neuron 1) perceives a periodic input signal.} $\sigma_1$ and $\sigma_2$ represent the strength of the coupling of neuron 2 to neuron 1, and of neuron 1 to neuron 2, respectively.}
\label{Fig12}
\end{figure}

The values of the parameters, $a=1.05$ and $\epsilon = 0.01$, are chosen such that, when $D=0$ and $\sigma_1=\sigma_2=0$, the neurons are in the excitable regime:  each neuron resides in a stable state (rest state) unless it is perturbed. If a strong enough perturbation occurs, the neuron leaves the rest state and after firing a spike, it returns to the rest state. Then, a refractory period follows during which another perturbation will not trigger a spike.

The equations are integrated, starting from random initial conditions, using the Euler-Maruyama method with an integration step of $dt = 10^{-3}$. The signal parameters, $a_0$ and $T$, and the coupling coefficients, $\sigma_1$ and $\sigma_2$, are varied within the ``subthreshold'' region of the parameter space: without noise the voltage-like variables $u_1$ and $u_2$ display only small oscillations [see Fig. 1(a)]. For each set of parameters, the voltage-like variable of the neuron that receives the signal, $u_1$, is analyzed and the ISI sequence is computed, $\{I_i; I_i= t_{i+1} - t_{i}\}$ with $t_i$ defined by the condition $u_1(t_i) = 0$ considering only the ascensions. 

To compute the mean ISI and the coefficient $R$ (see \textit{Methods}) time-series with a minimum number of 100 spikes are generated (as this is sufficient to estimate the mean values of the ISI distribution), while to compute the ordinal probabilities, time-series with at least 10000 spikes are generated. This is because a large number of ordinal patterns are needed in order to determine if their probabilities are consistent or not with the uniform distribution \cite{REI16_2}.

\subsection*{Methods}

The regularity of the ISI sequence is often characterized by the coefficient $R$~\cite{PIK97}:
\begin{equation}
R=\frac{\sqrt{\langle I^2 \rangle - \langle I \rangle^2}}{\langle I \rangle},
\end{equation}                                             
where $\langle I \rangle$ is the mean value of the ISI distribution. 

Correlations between ISIs are characterized by the serial correlation coefficients (SCCs):
\begin{equation}
C_j = \frac{\langle (I_i - \langle I \rangle)(I_{i-j} - \langle I \rangle)}{\langle I^2 \rangle - \langle I \rangle^2}
\label{eq:SCCs}
\end{equation}                                              
where $j$ is an integer number. 

SCCs are a standard tool to analyze spike trains \cite{neiman_pre_2005,andre_pre_2017}, however, they only capture linear correlations. In contrast, a symbolic methodology known as \textit{ordinal analysis} \cite{BAN02} has been demonstrated to be well suited for detecting nonlinear correlations in spike trains \cite{amigo_2013,REI16,pre_2009}. In this approach the actual ISI values $\{I_{1}, ..., I_i, ..., I_{N}\}$ are not taken into account, instead, their relative temporal ordering is considered. Ordinal analysis transforms a particular signal into symbols, which are known as ordinal patterns. Here, ordinal analysis is used to study the spike train of neuron 1: the ISI sequence $\{I_{1}, ..., I_i, ..., I_{N}\}$ is transformed into a sequence of ordinal patterns, which are defined by the relative order of $L$ consecutive ISI values. 

Once the length $L$ of the ordinal patterns is defined, for each interval $I_i$ the subsequent $L - 1$ intervals are considered and compared. The total number of possible order relations (i.e., ordinal patterns of length $L$) is then equal to the number of permutations $L!$. If we set $L =2$ we have only two patterns: 12 and 21 for $I_1 < I_2$ and $I_1 > I_2$, respectively, but if we set $L = 3$, we have 3! = 6 possible ordinal patterns, which are listed in Table \ref{t:table1}. For example, we consider the following sequence of intervals $\{4.9, 3.4, 3.3, 3.2, 5.0, ...\}$. The first value $I_1 = 4.9$, when compare with $I_2$ = 3.4 and $I_3$ = 3.3 leads to the ordinal pattern 210 since $I_1 > I_2 > I_3$. As well for $I_2$, since $I_2 > I_3 > I_4$. But for $I_3$ we have pattern 102 since $I_4 < I_3 < I_5$. 

\begin{table}[!ht]
\begin{adjustwidth}{-2.25in}{0in} 
\centering
 \captionof{table}{Ordinal patterns for $L = 3$}
\begin{tabular}{ ||c c||} 
\hline
 \textsc{Symbol} & \textsc{Relation} \\
 \hline\hline
012 & $I_3 > I_2 > I_1$ \\ 
021 & $I_2 > I_3 > I_1$\\
102 & $I_3 > I_1 > I_2$ \\
120 & $I_2 > I_1 > I_3$ \\
201 & $I_1 > I_3 > I_2$ \\
210 & $I_1 > I_2 > I_3$ \\
\hline
\end{tabular}
 \label{t:table1}
 \end{adjustwidth}
\end{table}

The symbolic sequence of ordinal patterns is computed using the function \texttt{perm} \underline{  } \texttt{indices}  defined in \cite{PAR12}. Then, the ordinal probabilities are estimated as $p_i=N_i/M$ where $N_i$ denotes the number of times the i-th pattern occurs in the sequence, and $M=\sum_{i=1}^{L!} N_i$ denotes the total number of patterns. If the patterns are equi-probable one can infer that there are no preferred order relations in the timing of the spikes. On the other hand, the presence of frequent (or infrequent) patterns will result into a non-uniform distribution of the ordinal patterns. A binomial test will be used to analyze the significance of preferred and infrequent patterns: if all the ordinal probabilities are within the interval $[p - 3\sigma, p + 3\sigma]$ (with $p = 1/L!$ and $\sigma = \sqrt{p(1-p)/M}$), the probabilities are consistent with the uniform distribution, else, there are significant deviations which reveal the presence of preferred and infrequent patterns. A main advantage of this method is that it is simple to implement and can be applied directly to the ISI sequence (no need to pre-process the data).

Here we use $L=3$, which allows to investigate order relations among three ISI (i.e., four consecutive spike times). This choice is motivated by the fact that the signal parameters and the coupling strengths are subthreshold, i.e., the firing activity of neuron 1 is driven by white noise (without noise, there are no spikes). Therefore, only short ISI correlations are expected in the spike train.


\section*{Acknowledgments}
This work was supported by Spanish MINECO (FIS2015-66503-C3-2-P) and the program ICREA ACADEMIA of Generalitat de Catalunya.
\nolinenumbers


%
%

\end{document}